\documentstyle[preprint,aps,prb,epsfig]{revtex}
\preprint{submitted to Appl. Phys. Lett. on April 24, 2003}

\begin{document}

\def\ws{{$\omega_s$}}
\def\wc{{$\omega_0$}}
\def\wp{{$\omega_0$+$\omega_s$}}
\def\wm{{$\omega_0$-$\omega_s$}}
\def\wpm{{$\omega_0 \pm \omega_s$}}

\def\twc{{$t(\omega_0)$}}
\def\twp{{$t(\omega_0$+$\omega_s)$}}
\def\twm{{$t(\omega_0$-$\omega_s)$}}
\def\twpm{{$t(\omega_0 \pm \omega_s)$}}

\def\ls{\lambda_s}

\draft
%\tighten

\title{High-contrast optical modulation by surface  acoustic waves}
\author{Srinivasan Krishnamurthy\footnote{Corresponding Author:
srini@aristotle.sri.com\\
Permanent address: SRI International, Menlo Park, CA 94025, USA}  and
P.~V.~Santos} 
\address{Paul-Drude-Institut f\"ur Festk\"orperelektronik,
        Hausvogteiplatz 5--7,
        10117 Berlin, Germany}
%\address{$^a$ Permanent address: SRI International, Menlo Park, CA
%94025, USA}
%
\date{\today}
\maketitle

%%%%%%%%%%%%%%%%%%%%%%%%%%%%%%%%%%%%%%%%%%%%%%%%%%%%%%%%%%%%%%
\begin{abstract}

Numerical calculations are employed to study the modulation of light
by surface acoustic waves (SAWs) in photonic band gap (PBG)
structures.  The on/off contrast ratio in a PBG switch based on an
optical cavity is determined as a function of the SAW-induced
dielectric modulation.  We show that these structures exhibit high
contrast ratios even for moderate acousto-optic coupling.

\end{abstract}

%%%%%%%%%%%%%%%%%%%%%%%%%%%%%%%%%%%%%%%%%%%%%%%%%%%%%%%%%%%%%%
%\begin{multicols}{2}

The interaction of light with acoustic waves in materials is of both
fundamental and technological interest. When the acoustic wave
propagates in the medium, it generates a strain field. Under
appropriate conditions, the associated periodic change in the
refractive index causes the light to undergo Bragg
reflection. Acoustic waves thus offer a convenient way to dynamically
modulate light propagation.  This effect has been used in a number of
applications including modulators, signal processors, tunable filters,
and beam deflectors.$^{1}$ Due to the weakness of the acousto-optic
(AO) interaction, however, many periods of the acoustic modulation are
normally required to increase the contrast between the transmitting
and reflecting states.  The use of surface acoustic waves (SAWs) in
the place of bulk waves has additional advantages.$^{2,3}$
High-frequency SAWs with large acoustic power densities can be easily
generated by inter-digital transducers (IDT) on piezoelectric
materials such as GaAs.  Since the fabrication of the IDTs employs
conventional semiconductor technologies, on-chip integration to active
devices becomes possible.

In this letter, we propose a design that combines SAWs with an optical
cavity inserted in an asymmetric Bragg mirror to modulate an incident
light beam through the AO effect.  When Bragg mirrors with
high-contrast dielectric materials are used, the band gap in the light
energy dispersion is large, while the 'defect' states introduced in
the gap by the cavity have a narrow energy distribution. Because of
the strong localization of the electromagnetic field, the AO
interaction becomes significantly enhanced in the cavity region, thus
resulting in high dielectric modulation.  Recently, an optical
modulator based on the optical Stark effect in a GaAs cavity has been
demonstrated$^{4}$ to yield a ratio of transmitted intensity in the
'on' state to that in the 'off' state, known as the contrast ratio
(CR), of 5. We show that higher CRs can also be achieved with modest
acoustic power in a SAW-based optical modulator.

In photonic band gap (PBG) crystals, such as Bragg mirrors, a
structural modulation is used to create a periodic change in the
refractive index (and, consequently, in the dielectric
function).$^{5}$ This periodicity is comparable to the light
wavelength.  When a SAW with a wavelength $\ls$ much longer than the
periodicity of the underlying PBG structure propagates across the
structure, an additional time-dependent periodicity is superimposed on
its dielectric function, as shown in Fig.~1.  The solution to
Maxwell's equations has to include the effects of both the short- and
the long-range periodicities.  We have extended the mathematically
rigorous transfer matrix (TM) method$^{6-8}$ to the case where the
unit cell in the direction of light propagation is a multiple of $\ls$
and the dielectric function varies with the SAW frequency \ws.  When
light of frequency \wc$>>$\ws~ impinges on the PBG material, the
transmission and reflection take place through channels of frequencies
(\wc $\pm m$ \ws) with $m=0,1,2,\dots$.$^9$ In the present
calculations, we neglected the modes with $m\ge2$. The size of the
transmission and reflection matrices in the TM approach becomes, in
this case, three times as large as that for a bare PBG structure
(i.e., in the absence of a SAW). Care has been taken to remove
numerical instability arising on account of the increase in the number
of forward steps {\it and} in the matrix dimensions. The method has
been found to be stable even for structures containing as much as 300
PBG unit cells in a SAW wavelength.  The details of this calculational
procedure will be published elsewhere.$^{10}$

For ease of fabrication, we considered thin and planar structures,
which can be grown with the current molecular-beam-epitaxy (MBE)
technology.  The composition and thickness of the layers as well as
the total size of the layer stack were varied so as to enhance the
interaction between light of wavelength of 940 nm and a SAW with
$\ls=5.6~\mu$m.  The electric field concentration in the cavity is
crucial for the enhancement of AO interaction. We found that the
modulation is the largest when the cavity is placed close to, but not
exactly at the center of the Bragg mirror structure. One possible
structure is shown in Fig.~1. It is a 40-period asymmetric Bragg
structure with a GaAs cavity near the center. The first 21 periods are
alternating AlAs and GaAs layers with thicknesses of 76.9~nm and
65.5~nm, respectively, followed by the 142.4~nm-thick GaAs cavity and
the second mirror, which has 18 periods with the same composition as
the first Bragg mirror. This structure exhibits a forbidden gap of
117~nm (from 865~nm to 983~nm), as shown in the spectrum of its
transmission coefficient in Fig.~2. The cavity states are located near
940~nm and have a full width at half maximum (FWHM) of about
0.5~nm. Although this design is for operation near 940 nm, the
structure can be easily scaled for other operation frequencies.

In our calculations, we assumed that the SAW creates a sinusoidal
modulation of the dielectric constant, as shown in Fig.~1, with a peak
value for the relative modulation of the dielectric function $\Delta
\epsilon$/$\epsilon = 6 \times 10^{-4}$. From calculations of the SAW
strain field, we determined that these dielectric modulation levels
can be generated in GaAs/AlAs multilayers by conventional IDTs
excited with a 500~MHz radio-frequency power of 20~mW.  In Fig.~3, the
calculated transmission coefficients through the central channel
[\twc, dots] and side channels [\twpm, squares] are compared with
that obtained in the absence of a SAW [$t_0$, open circles].  The
inter- and the intra-channel scattering modes induced by the SAW have
negligible intensities except at frequencies very close to the cavity
states, where the AO interaction is strong.  Although the transmission
occurs through these three frequencies, the channels are separated
only by a very small frequency \ws$<<$\wc. In experiments, the
observed transmission coefficient will thus be the sum of the
contributions from all three channels. Hence, the CR is defined as the
ratio of $t_0$ to [\twc+\twp+\twm] and plotted as a dashed
line in Fig.~3. A maximum CR of about 4 can be obtained.  The
dielectric modulation $\Delta\epsilon$, which depends on the square
root of the acoustic power,$^{11-13}$ can be further increased by
improving the IDT design.  Previous calculations for a specific
design$^{13}$ within the GaAs/(Al,Ga)As system show that much larger
$\Delta\epsilon /\epsilon$ ratios (of up to $2.3\%$) are
possible.  To investigate the operation of the switch under theses
conditions, we show in Fig.~4 the calculated CRs for various values of
$\Delta\epsilon$/$\epsilon$.  CR increases almost quadratically with
$\Delta\epsilon /\epsilon$, reaching values as high as 300 for
$\Delta\epsilon /\epsilon= 0.6\%$ (corresponding to a relative
refractive index change of 0.3\%).

Layer size fluctuations inherently present in MBE growth may affect
the properties of the Bragg mirrors and deteriorate the contrast
ratio.  This effect was modeled by assuming that the dielectric
constants fluctuated randomly within 5\% of their original values. We
found that the highest CR decreases from 300 to 240. In addition, the
central frequency may shift as the cavity thickness changes.

Although the structure proposed here can be grown by the MBE method,
an unconventional placement of the IDT for SAW generation may be
required.  Since the dimensions of IDT for high frequency SAWs
($>$1~GHz) can be significantly reduced below a mm, we suggest its
placement on the {\it side} of the substrate, as shown in Fig.~5. In
this configuration, both the SAW and the light beams travel along the
direction perpendicular to the layer stack.  Recent developments of
electron-beam lithography and of imprint techniques$^{14}$ for IDT
fabrication may make such an arrangement possible.  Another
possibility is to generate the SAW on a highly piezoelectric material,
such as LiNbO$_3$ or ZnO, and then to couple it to the device
structure on GaAs employing wafer bonding techniques.

While the cavity design leads to large CR values, it has a limited
frequency tunability, because of the $\delta$-function-like
distribution of defect modes. A wider tunability range can be achieved
if the SAW interacts with propagating light modes lying outside the
forbidden gap of the Bragg mirror.  The strength of the AO
interaction, however, is considerably weaker for these extended
modes.$^{9,15}$ While operating near the zone edge might improve the
AO coupling (because of the reduced light group velocity), longer SAW
wavelengths and, consequently, longer interaction paths will be needed
to ensure wave vector conservation.  For shorter switches, such as the
one based on the one-dimensional Bragg stack, it is clear that one has
to work in a band with extremely small dispersion in order to achieve
large light modulation efficiencies.

The major advantages of the optical switch proposed here are the
potentially very high on/off CRs, sub-$\mu$s switching times, and
extremely small sizes.  The frequency tunability (selection of \wc) is
limited to a sub-nm range defined by the cavity states. If more cavity
states are introduced to increase the energy width and the tunability
range, the strength of the AO interaction will be reduced.  The latter
may be compensated by increasing the number of layer stacks.  The
growth of high-quality Bragg mirrors by MBE, however, is restricted to
a maximum stack thickness of a few microns.

In conclusion, we have used an extended version of the TM approach to
calculate the changes in the transmission spectra induced by SAWs in
PBG crystals. We show that a large modulation of light can be achieved
in a one-dimensional Bragg stack with a cavity.  If the acoustic power
can be increased to cause a 0.30\% change in the refractive index near
the cavity in the PBG structure, CRs of the order of a few hundred can
be achieved through the AO effect.

The original TM code, written by A.J. Ward and J.B. Pendry, is from
the Computer Physics Communication International Program Library.  We
thank Prof. J.B. Pendry for providing additional material to
understand their computer code, Dr. A. Reynolds for a discussion on
numerical stability, and Prof. H. Grahn for comments on the
manuscript.  Financial support (for S.K) from the Alexander von
Humboldt Foundation, Germany, and from the Deutsche
Forschungsgemeinschaft (for P.V.S., project No. SA598/3-1) is
gratefully acknowledged.

%%%%%%%%%%%%%%%%%%%%%%%%%%%%%%%%%%%%%%%%%%%%%%%%%%%%%%%%%%%%%%%%%%%%%%%%

%%%%%%%%%%%%%%%%%%%%%%%%%%%%%%%%%%%%%%%%%%%%%%%%%%%%%%%%%%%%%%%%%%%%%%%%
\setcounter{figure}{0}

%Fig.~1
\begin{figure}
\caption{Dielectric constant $\epsilon$ (solid line)  of the 40 layer asymmetric
  Bragg structure with the modulation  $\Delta \epsilon$ (dashed line)  superimposed
by a SAW.}
\label{fig1}
\end{figure}
%===========================================================================
%Fig.~2
\begin{figure}
\caption{Calculated transmission coefficient  for the structure
in Fig.~1 in the absence of a SAW.}
\label{fig2}
\end{figure}
%===========================================================================
%Fig.~3
\begin{figure}
\caption{ Calculated transmission coefficient for the structure given
in Fig.~1 with and without a SAW. The dotted  line shows the
contrast ratio CR.}
\label{fig3}
\end{figure}
%===========================================================================
%Fig.~4
\begin{figure}
\caption{Expected contrast ratio CR as a function
of the dielectric modulation $\Delta\epsilon/\epsilon$.} 
\label{fig4}
\end{figure}
%===========================================================================
%Fig.~5
\begin{figure}
\caption{Design of one-dimensional photonic stack with an IDT
deposited on a cleaved edge of the substrate. }
\label{fig5}
\end{figure}
%===========================================================================
%\end{multicols}
%\end{document}
\setcounter{figure}{0}
\newpage
%Fig.~1
\begin{figure}
\centerline{\epsfig{file=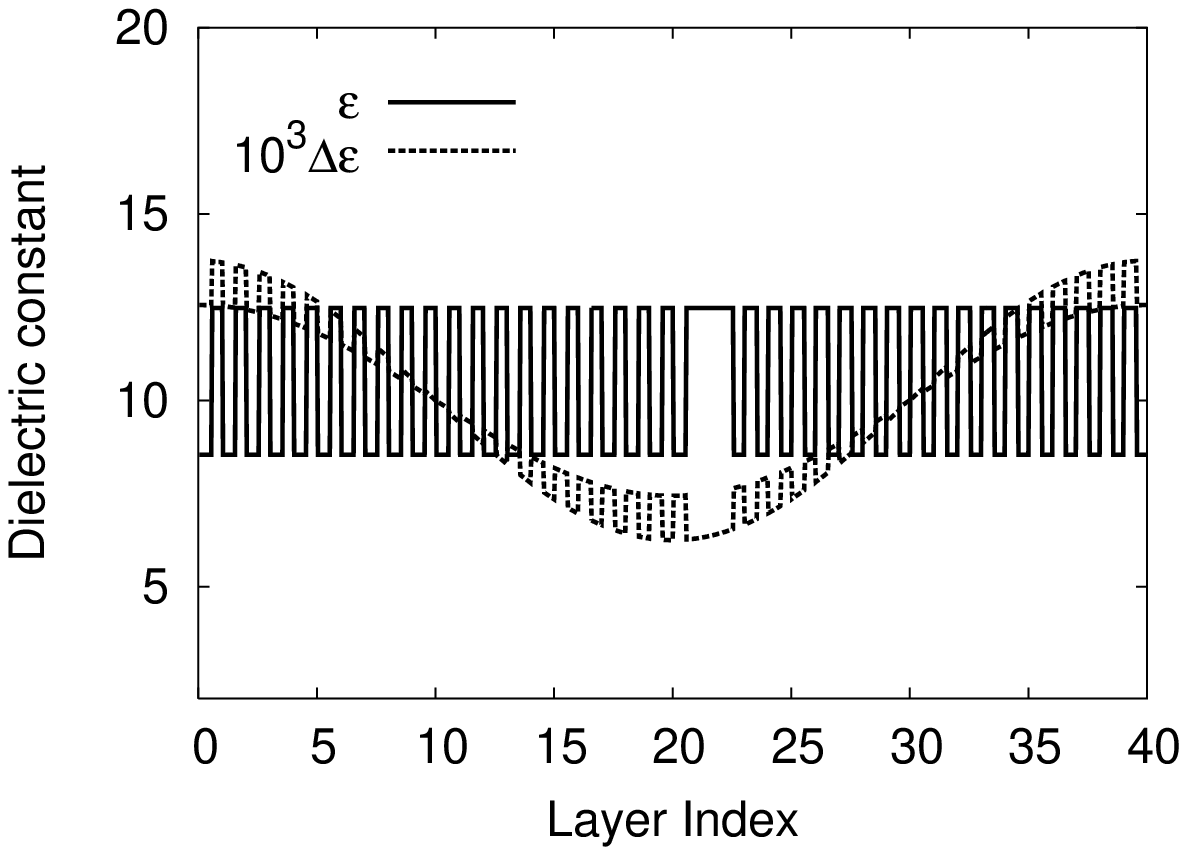,width=8.0cm}}
\caption{ Krishnamurthy {\it et al.}}
\end{figure}
%===========================================================================
\newpage
%Fig.~2
\begin{figure}
\centerline{\epsfig{file=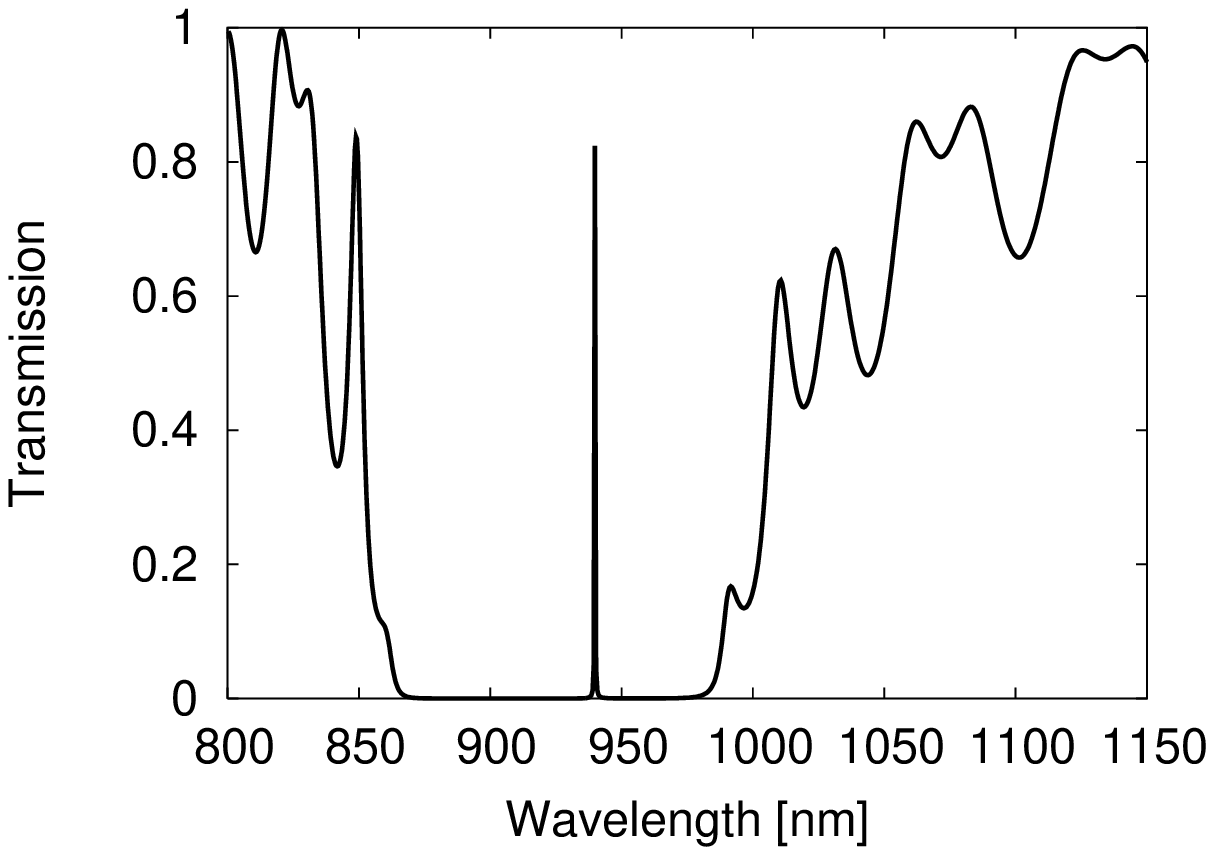,width=8.0cm}}
\caption{Krishnamurthy {\it et al.}}
\end{figure}
%===========================================================================
\newpage
%Fig.~3
\begin{figure}
\centerline{\epsfig{file=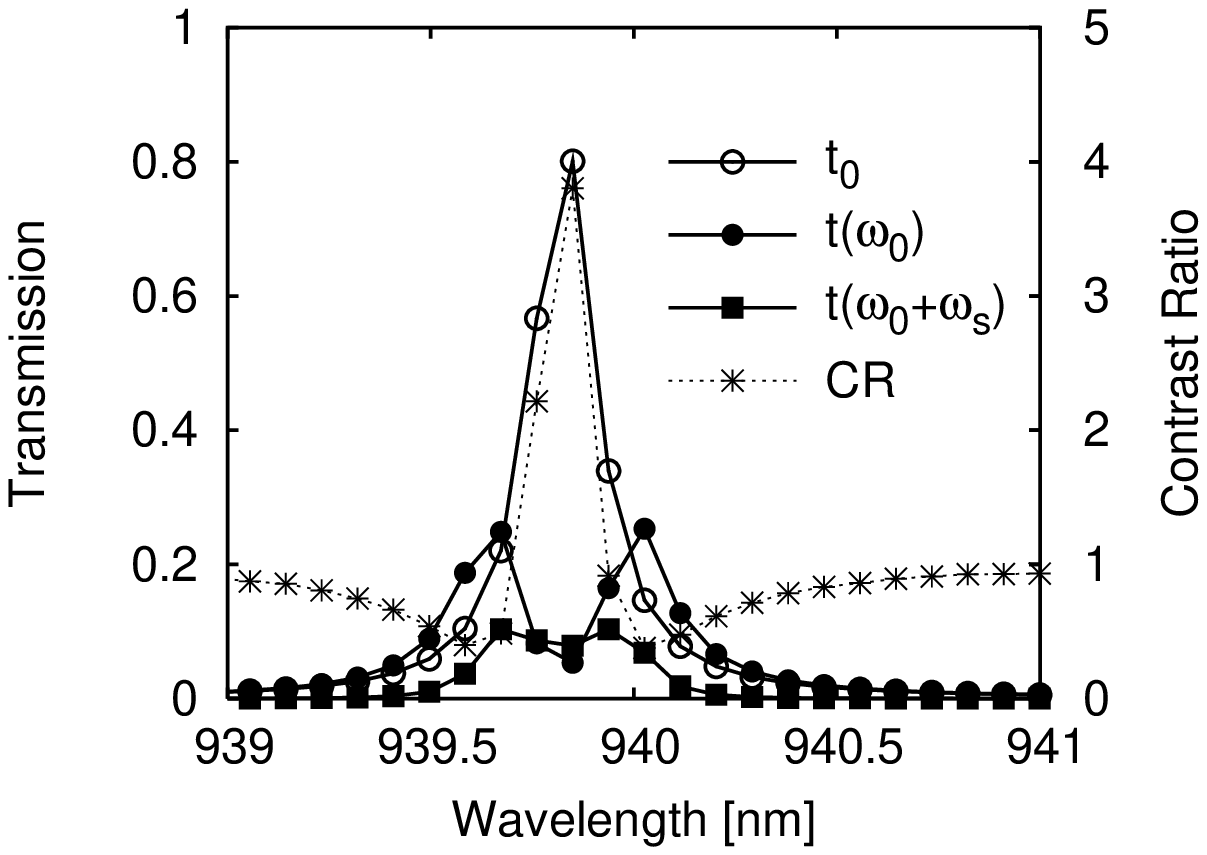,width=8.0cm}}
\caption{ Krishnamurthy {\it et al.}}
\end{figure}
%===========================================================================
\newpage
%Fig.~4
\begin{figure}
\centerline{\epsfig{file=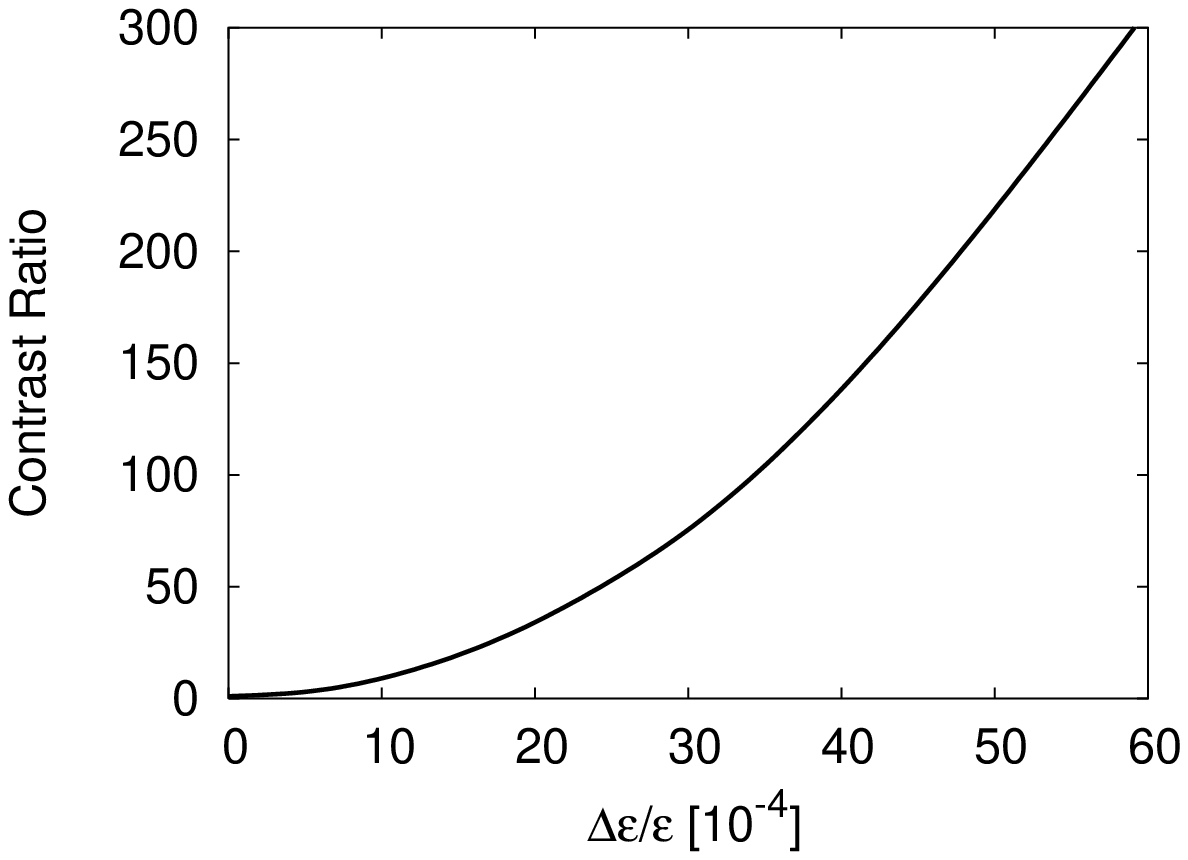,width=8.0cm}}
\caption{ Krishnamurthy {\it et al.}}
\end{figure}
%===========================================================================
\newpage
%Fig.~5
\begin{figure}
\centerline{\epsfig{file=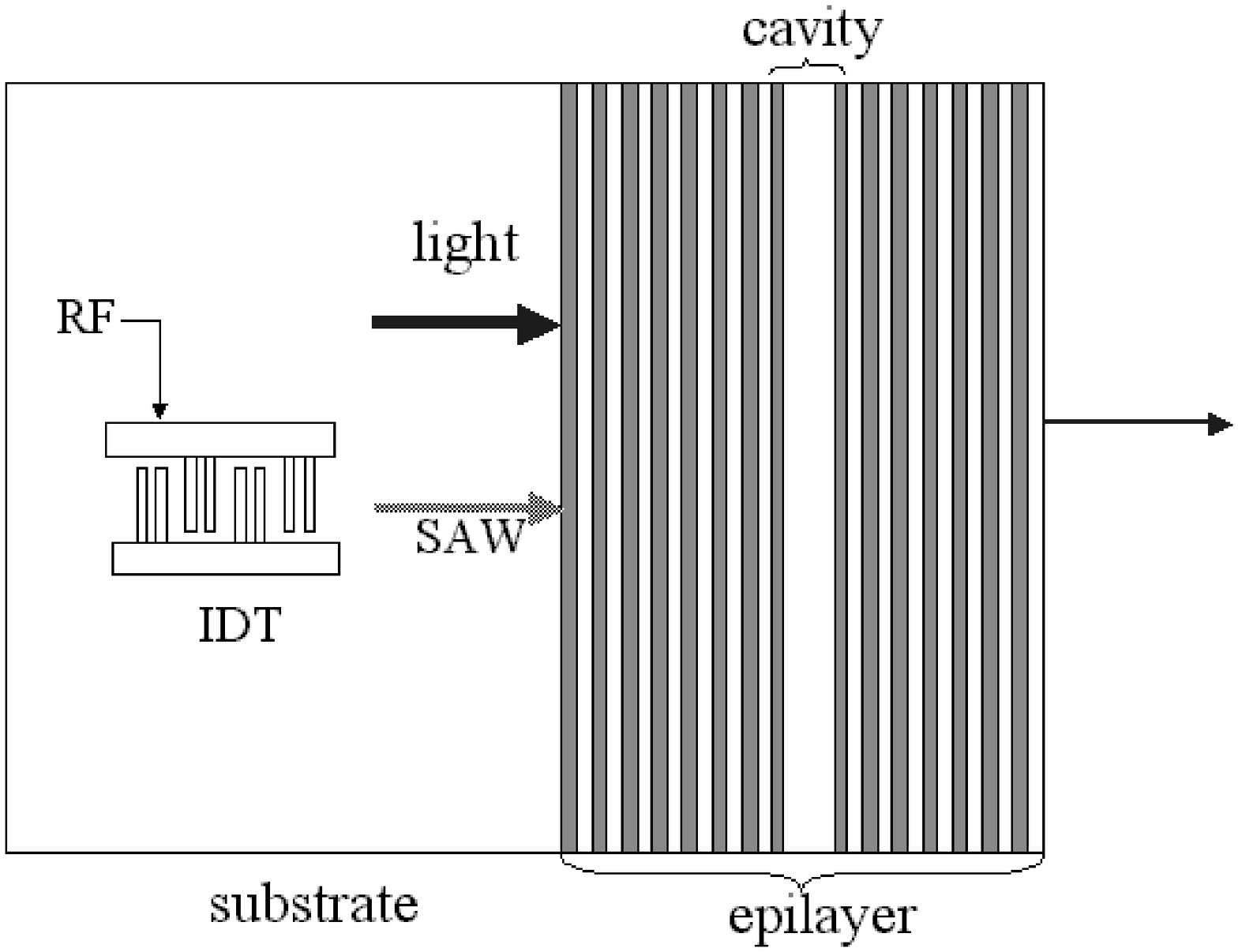,width=8.0cm}}
\caption{ Krishnamurthy {\it et al.}}
\end{figure}
%===========================================================================


\begin{references}

\bibitem{1} For example, N.~Goto and Y.~Miyazaki, Jpn. J. Appl. Phys.,
Part 1 {\bf 37}, 2947 (1998); and references cited therein.

\bibitem{2} P.~Hess, Phys. Today, {\bf 55}(3), 42 (2002).

\bibitem{3} C.W.~Ruppel, R.~Dill, A.~Fischerauer, G.~Fischerauer,
W.~Gawlik, J.~Machui, F.~Muller, L.~Reindly, W.~Ruile, G.~Scholl,
I.~Schropp, and K.C.~Wagner, IEEE Trans. Ultra. Ferro. and
Freq. Control {\bf 40}, 438 (1993).

\bibitem{4} S.~Sanchez, C.~De Matos, and M.~Pugnet, Appl. Phys. Lett.
{\bf 78}, 3779 (2001).

\bibitem{5} J.D.~Joannopoulos, R.D.~Meade, and J.N.~Winn, {\it
Photonic Crystals},  (Princeton University Press, Princeton, 1995).

\bibitem{6} J.B.~Pendry, J. Mod. Optics {\bf 41}, 209 (1994).

\bibitem{7} J.B.~Pendry, J. Phys. Cond. Matter {\bf 8}, 1085
(1996). 

\bibitem{8} A.J.~Ward and J.B.~Pendry, J. Mod. Optics {\bf 43}, 773
(1996).

\bibitem{9} P.V.~Santos, J. Appl. Phys. {\bf 89}, 5060 (2001).

\bibitem{10} S.~Krishnamurthy and P.V.~Santos, (in preparation). 

\bibitem{11} S.~Datta, {\it Surface Acoustic Waves},
(Prentice-Hall, Englewood Cliffs, 1986).

\bibitem{12} S.~Swierkowski, T.~van Duzer, and C.W.~Turner, IEEE
Trans. Son. Ultrason. {\bf SU-20}, 260 (1973).

\bibitem{13} F.C.~Jain and K.K.~Bhattacharjee, IEEE Photonics
Tech. Lett. {\bf 1} 307 (1989).

\bibitem{14} S.Y.~Chou, P.R.~Krauss, and P.J.~Renstrom,
Appl. Phys. Lett. {\bf 67}, 3114 (1995).

\bibitem{15} W.F.~Liu, P.St.J.~Russell, and L.~Dong, Opt. Lett. {\bf 22},
1515 (1997).

\end{references}
\end{document}